\documentclass{moriond}

\def\Journal#1#2#3#4{{#1} {\bf #2}, #3 (#4)}

\def\PLB{{\em Phys. Lett.}  B}

\usepackage{lineno}

\def\be{\begin{equation}}
\def\ee{\end{equation}}
\def\bea{\begin{eqnarray}}
\def\eea{\end{eqnarray}}

\usepackage{xspace}
\newcommand{\T}{\ensuremath{\mathrm{t}}} 
\newcommand{\Tb}{\ensuremath{\bar{\mathrm{t}}}} 
\newcommand{\ttbar}{\T\Tb\xspace} 
\newcommand{\ttbarN}{\T\Tb} 
\newcommand{\ttZ}{\ttbarN Z\xspace} 
\newcommand{\ttW}{\ttbarN W\xspace} 
\newcommand{\tW}{\ensuremath{\mathrm{tW}}\xspace}
\newcommand{\tZq}{\ensuremath{\mathrm{tZq}}\xspace}

\begin{document}
\vspace*{4cm}
\title{ASSOCIATED TOP PRODUCTION IN ATLAS AND CMS}

\author{ SERGIO S\'ANCHEZ CRUZ \\ on behalf of the ATLAS and CMS Collaborations}

\address{Facultad de Ciencias, 18 Federico Garc\'ia Lorca, \\
Oviedo (Asturias) Spain}

\newcommand{\Photo}{}

\maketitle

\abstracts{
A set of measurements of top quark pair and single top quark production in association with standard model (SM) bosons performed by the CMS and ATLAS Collaborations is presented. The consistency of these results among themselves and with the SM prediction is discussed. The latest results on four top quark production are also shown.
}

\section{Introduction}

The study of processes that involve the production of top quarks in association with the vector bosons is an important test of the consistency of the SM of particle physics. In these processes top quarks are either produced singly (\tZq, \tW) or in pairs (\ttZ, \ttW), and their production cross-sections in pp collisions are several orders of magnitude below  that of \ttbar production or Drell-Yan, characteristic of hadron colliders at the CERN LHC energy scale. Nevertheless the increased integrated luminosity delivered by the LHC to the ATLAS~\cite{exp_atlas} and CMS\cite{exp_cms} experiments makes the study of these processes feasible. This article reports on measurements of these processes in pp collisions at $\sqrt{s} = 13$ TeV recorded by the ATLAS and CMS experiments.

Measurements of \ttZ and \tZq provide a direct probe of the coupling of the top quark to the Z boson, and \tZq can also be sensitive to the WWZ coupling. The study of the \tW process is interesting because its sensitivity to the b quark PDF and its interference with \ttbar production at NLO. Therefore tW production tests interesting aspects of QCD. \ttW is, as the rest of the processes, an important background in searches for new physics and precision SM measurements.

\begin{figure}
	\includegraphics[width = 0.33\textwidth]{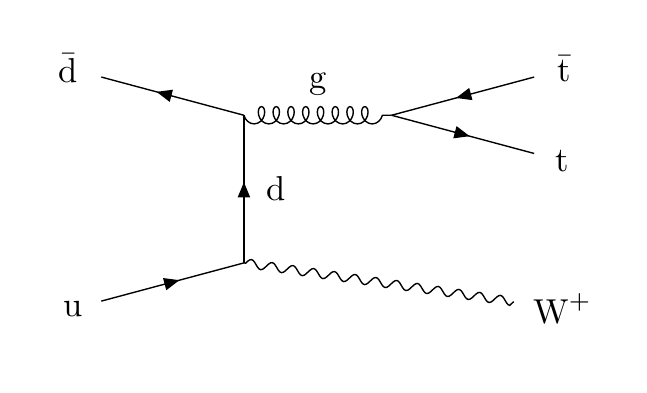}
	\includegraphics[width = 0.33\textwidth]{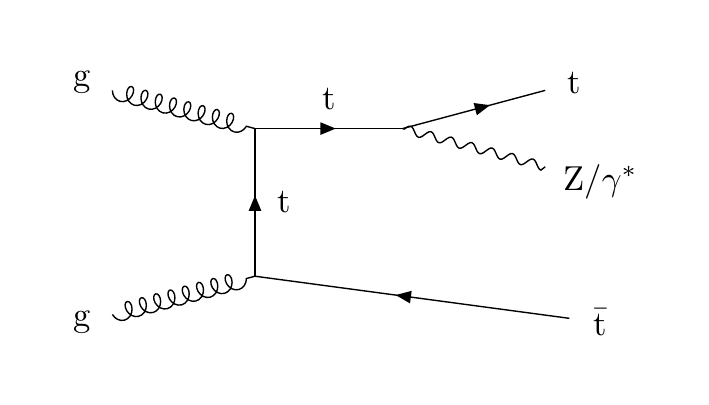}
	\includegraphics[width = 0.2 \textwidth]{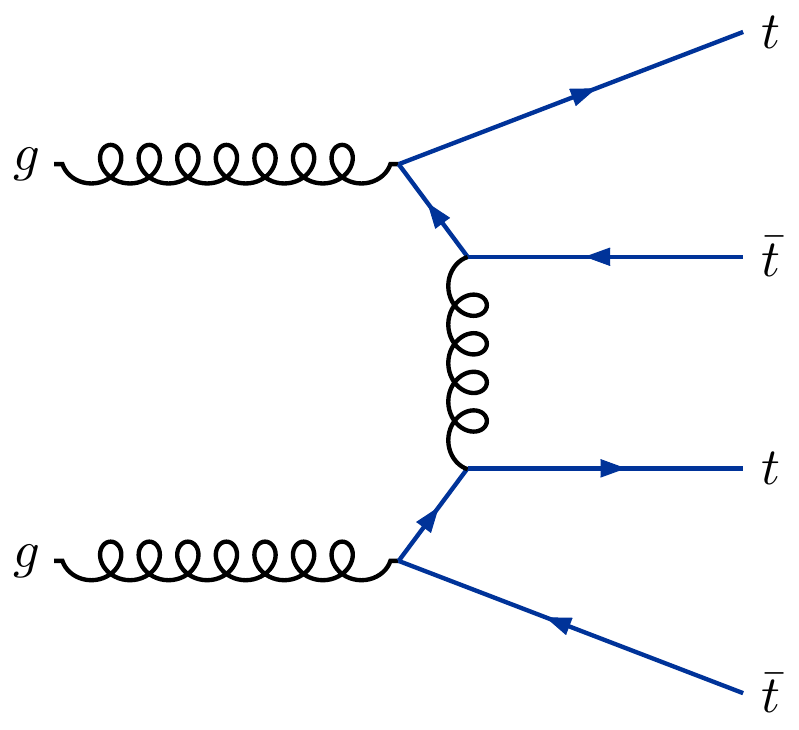}
	\begin{center}
	\includegraphics[width = 0.33\textwidth]{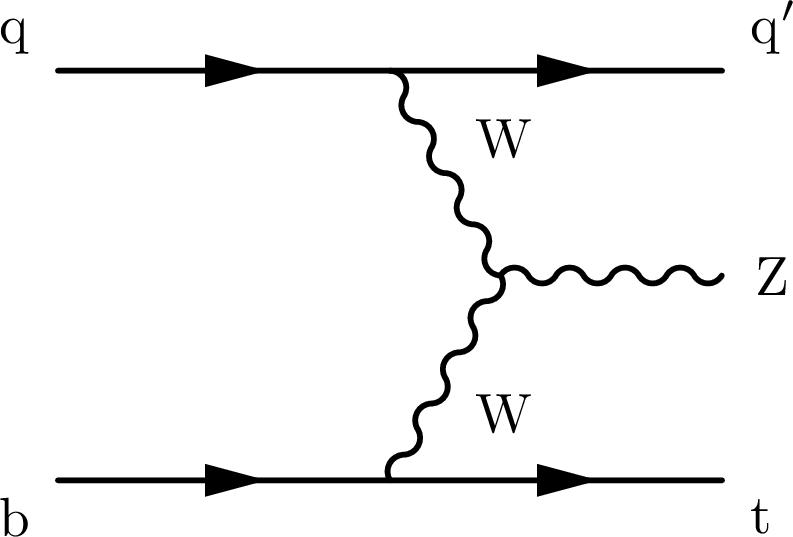}
	\includegraphics[width = 0.33 \textwidth]{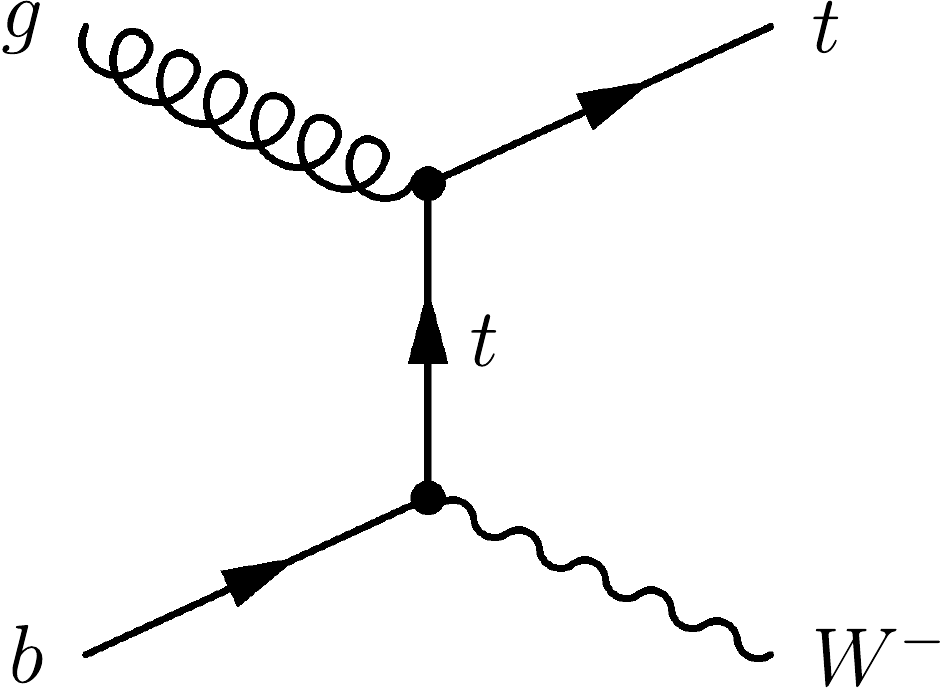}
	\end{center}

	\caption{Leading order diagrams for the processes studied in this document. Top row shows \ttW (left), \ttZ (center) and \ttbarN\ttbar (right), and bottom row shows \tZq (left), \tW (right).}
	\label{fig:diagram}
\end{figure}

Another interesting process is \ttbarN\ttbar production. Its small cross-section makes very challenging its study at the LHC, however its measurement would provide a valuable test of higher-order perturbative QCD calculations. Moreover, it is sensitive to the Yukawa coupling of the Higgs boson to the top quark, $y_t$.  Figure~\ref{fig:diagram} shows leading order diagram for the 5 processes.

\section{Associated production of a \ttbar pair with a vector boson}

The \ttW and \ttZ production cross-section measurements~\cite{cms_ttv,atlas_ttv} are performed in events in which at least one of the W bosons originated in the top quark decay further decays into a lepton (electron or muon), and the associated W and Z boson decays to a lepton and a neutrino or two leptons, respectively.

\begin{figure}
	\includegraphics[width=0.6\textwidth]{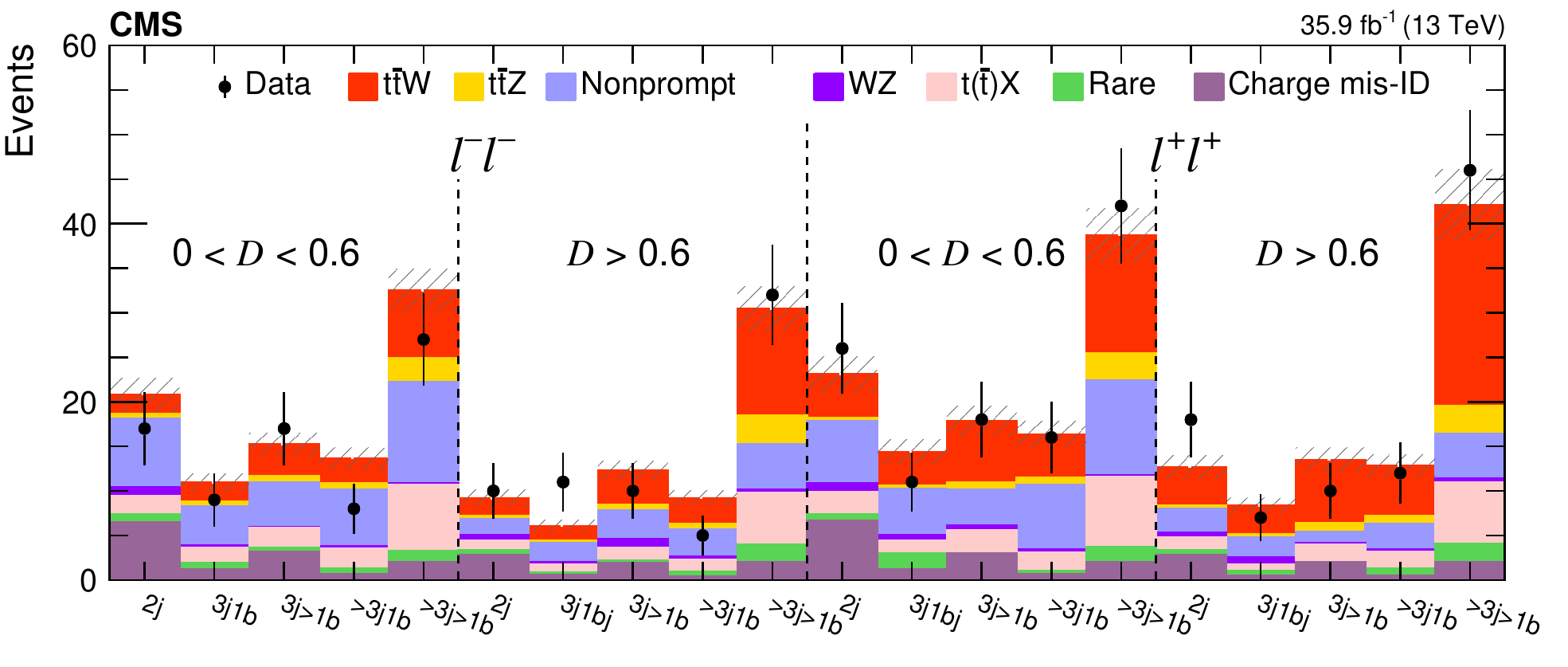}
	\includegraphics[width=0.4\textwidth]{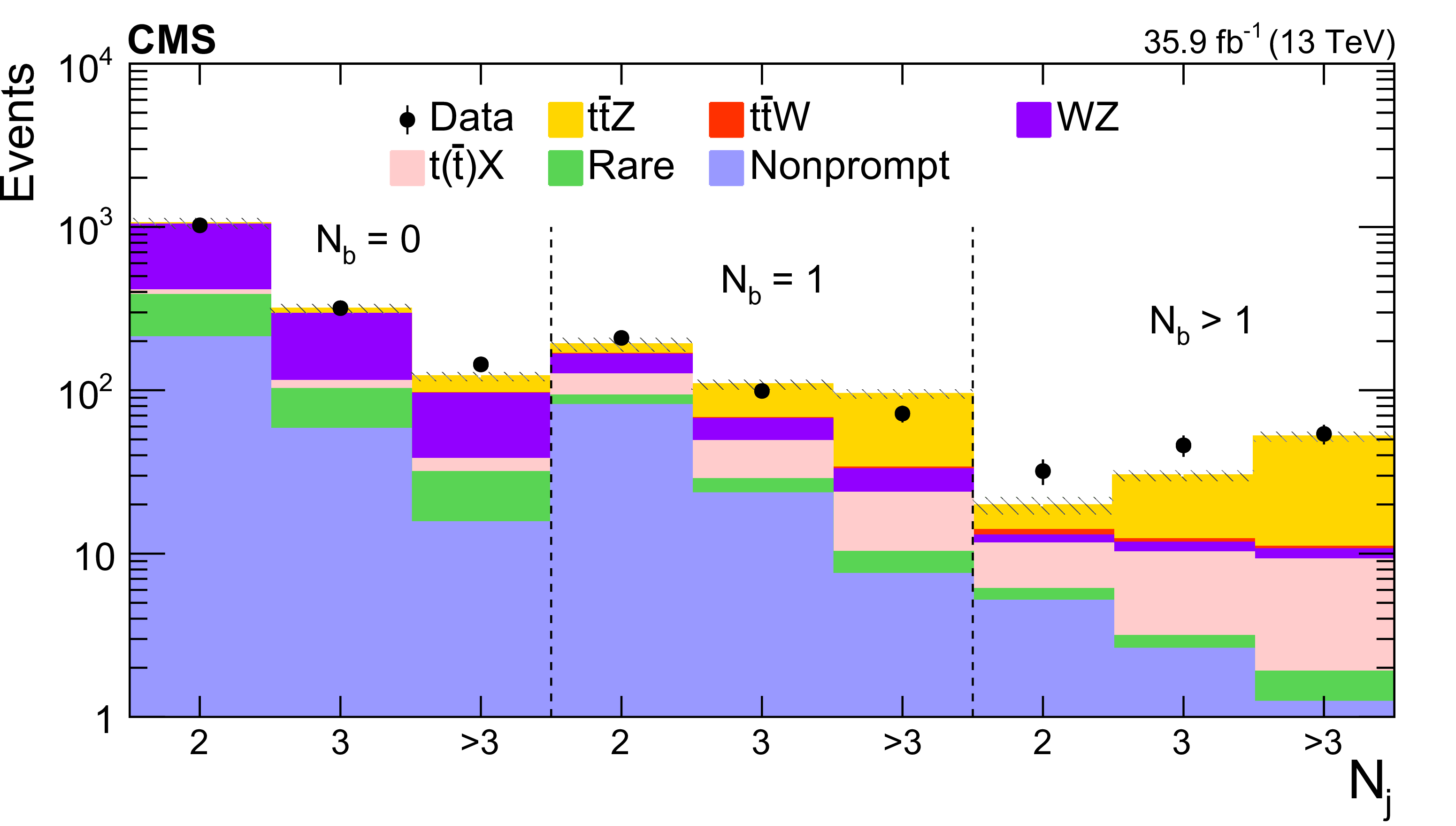}
	\caption{Event yield in measurement regions from which the signals are extracted, after the fit is performed~\protect\cite{cms_ttv}.}
	\label{fig:srs_ttv_cms}
\end{figure}

The production of \ttW is measured by CMS in events with two same-sign leptons, in order to discriminate most of the background, and utilizes multivariate analysis (MVA) techniques in order to separate signal from the remaining backgrounds, and to define a control and signal region. 20 independent measurement regions are built in order to extract the cross-section depending on the value of the MVA score, the charge of the leptons, the number of jets and number of b-tagged jets, as shown in fig~\ref{fig:srs_ttv_cms}. The measurement of ttZ is performed in events with three or four leptons containing a Z boson candidate, and measurement regions are defined depending on the number of jets and b-tagged jets. Signals are extracted by performing a maximum likelihood fit to the event yield in the measurement region defined.

\begin{figure}
		\centering
		\includegraphics[height=5cm]{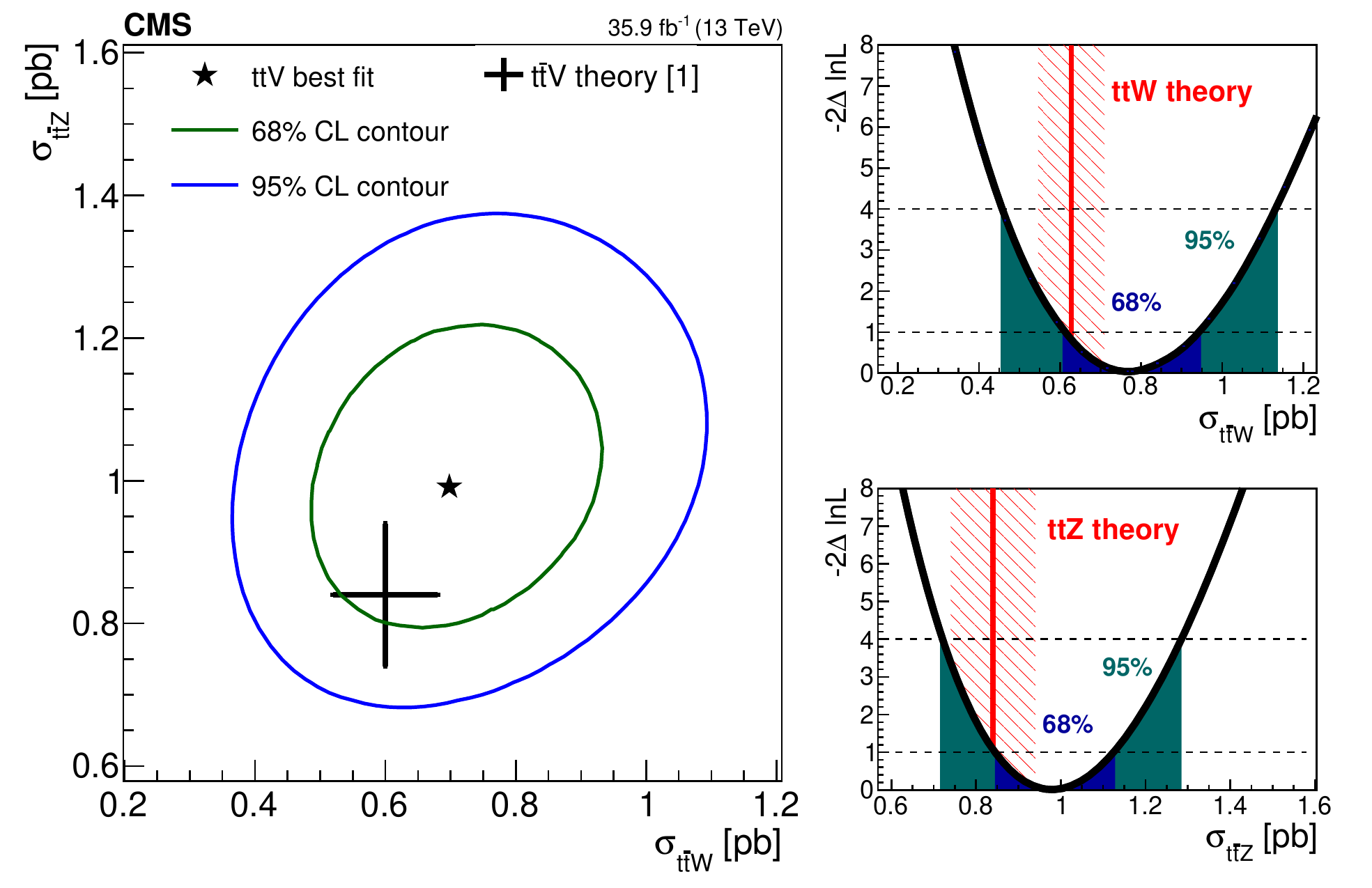}
		\includegraphics[height=5cm]{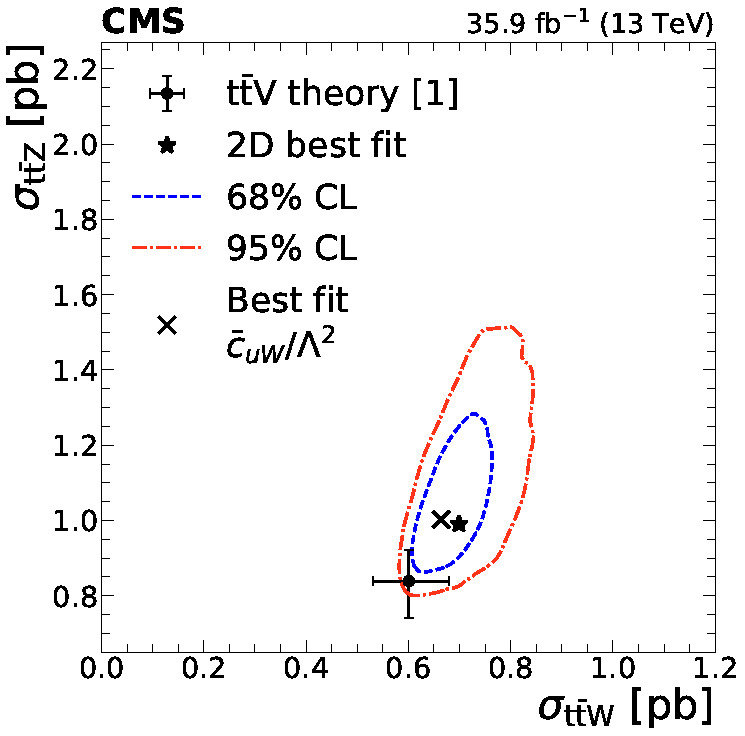}
		\caption{Results for measurements of the \ttW and \ttZ production in CMS. Left shows the confidence level contours for the simultaneous measurement of the two. Center shows the likelihood scans for the separated extraction of the two processes. Right shows the cross-section corresponded to the best fit of $\bar{c}_{uV}/\Lambda^2$~\protect\cite{cms_ttv}.}
		\label{fig:cms_results_ttv}

\end{figure}

Main backgrounds for both analyses are events containing non-prompt leptons, which are not produced in prompt decays of W or Z bosons and which are estimated in a data-driven fashion, and WZ production, that is estimated using simulations and validated in data control regions.

\ttZ and \ttW are determined  simultaneously and found to be consistent with SM prediction within one standard deviation, as shown in fig~\ref{fig:cms_results_ttv}. The result described corresponds to more than 5 $\sigma$ evidence of the two processes. This result allows to put stringent limit on new physics models parameterized by the Wilson coefficients of several dimension 6 operators~\cite{wilson_feyn}.

\begin{figure}
	\centering
	\includegraphics[height = 4.5cm]{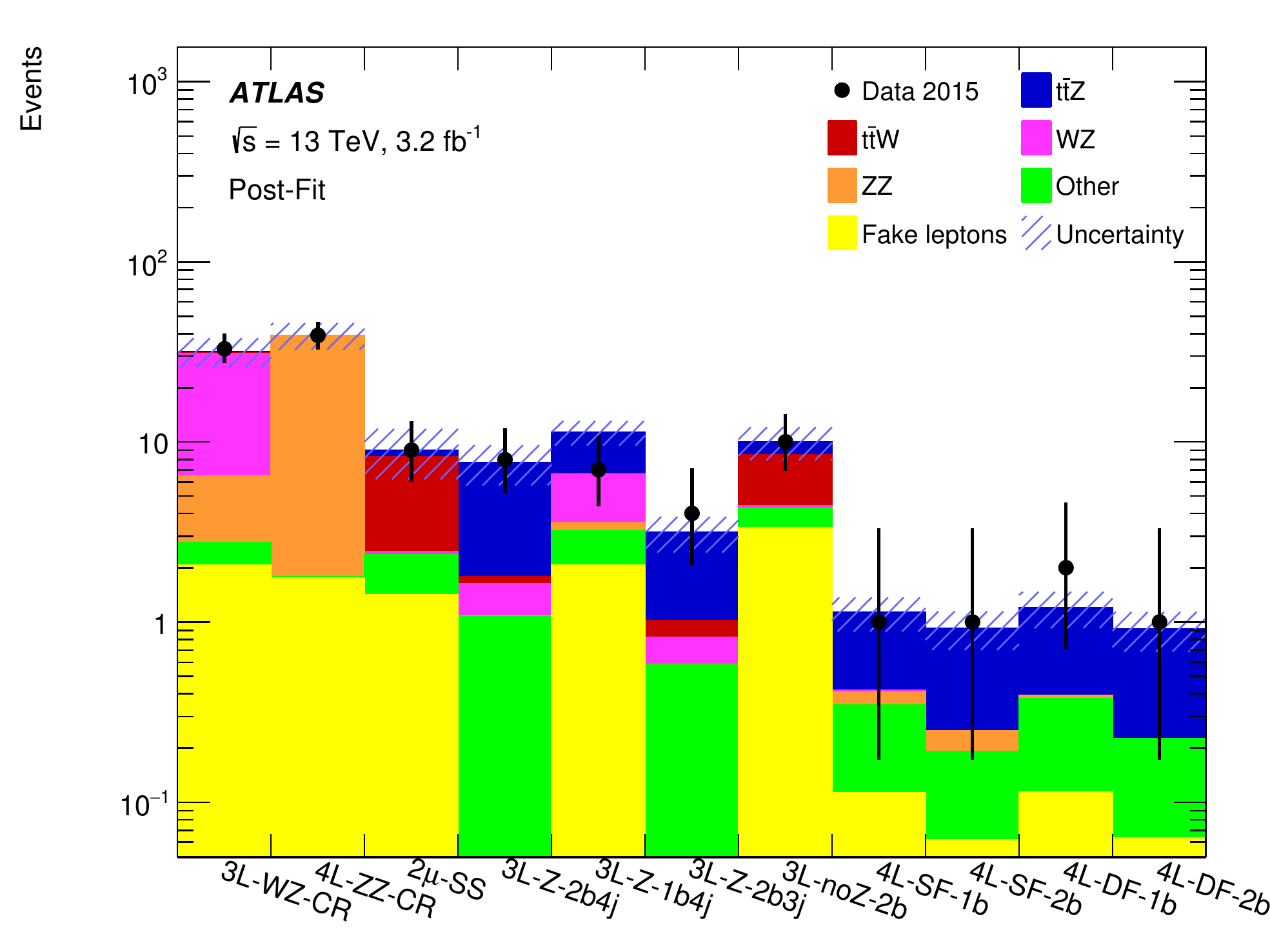}
	\includegraphics[height = 4.5cm]{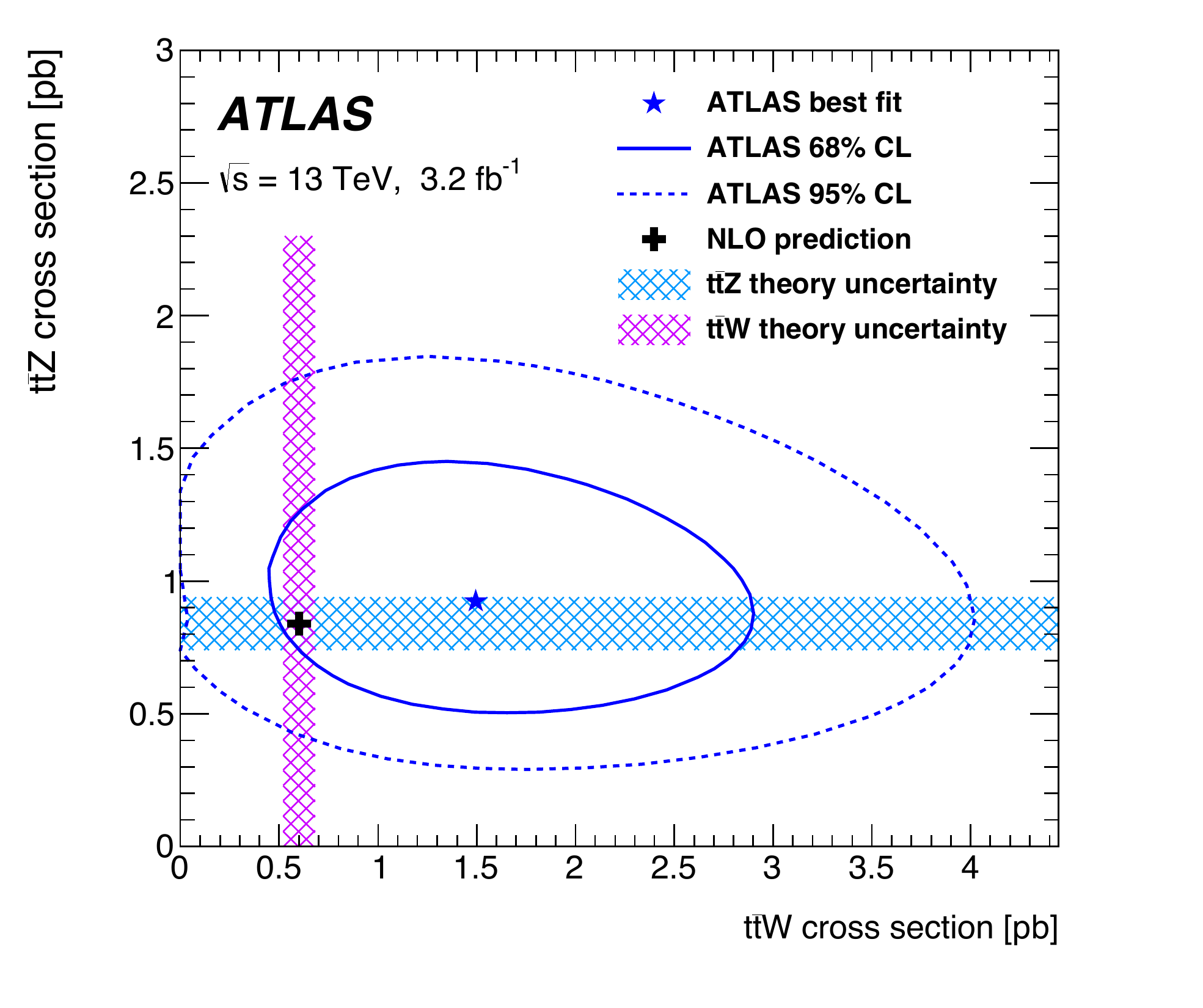}
	\caption{Event yield in measurement and control regions for the measurement of \ttW and \ttZ production (left), and confidence level contours for the simultaneous measurement of the two processes (right)~\protect\cite{atlas_ttv}.}
	\label{fig:srs_ttv_atlas}
\end{figure}

Another measurement performed by ATLAS targets equivalent topologies~\cite{atlas_ttv}. \ttW is measured in events with two same-sign muons and three leptons, while \ttZ is measured in events with three or four leptons. Several measurement and control regions are defined, depending on the topology of the event. They are summarized in figure~\ref{fig:srs_ttv_atlas}.

Being performed in a smaller dataset, the result is still limited by the statistical uncertainty of the collected data, and provides 3.9 $\sigma$ evidence for \ttZ production and 2.2 $\sigma$ evidence for \ttW production.

\section{\ttbarN\ttbar production}
\begin{figure}
	\includegraphics[height=4.5cm]{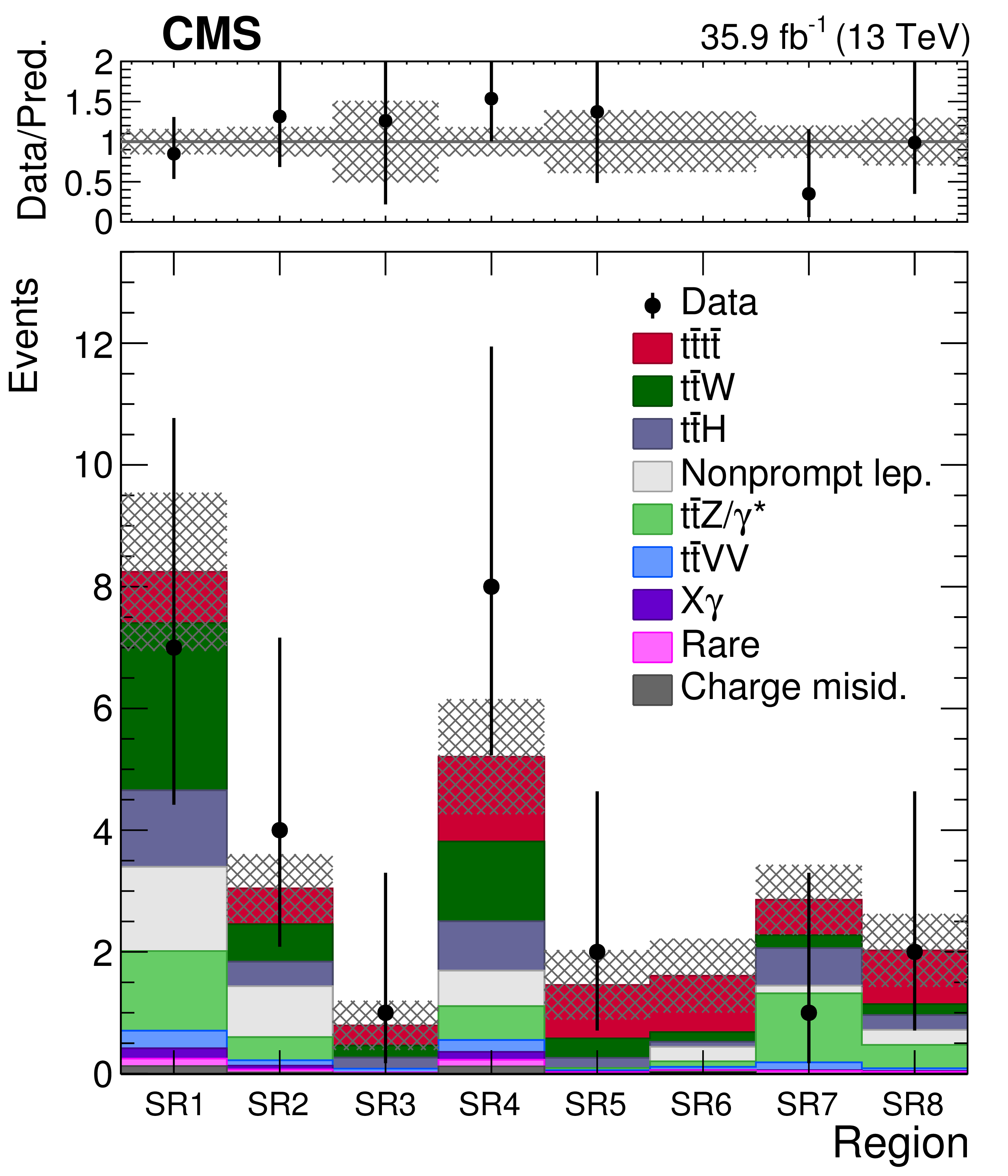}
	\includegraphics[height=4.5cm]{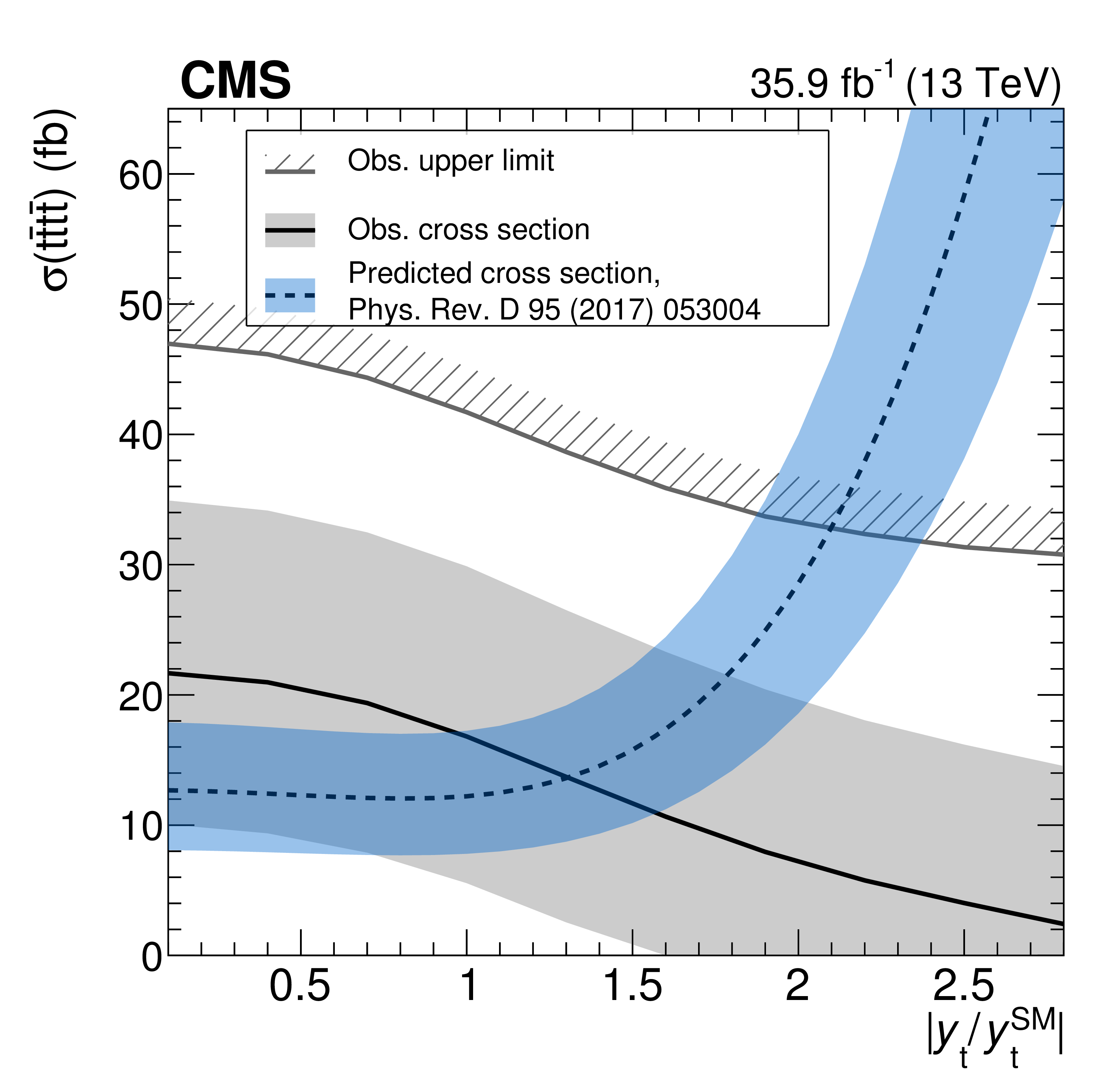}
	\includegraphics[height=4.5cm]{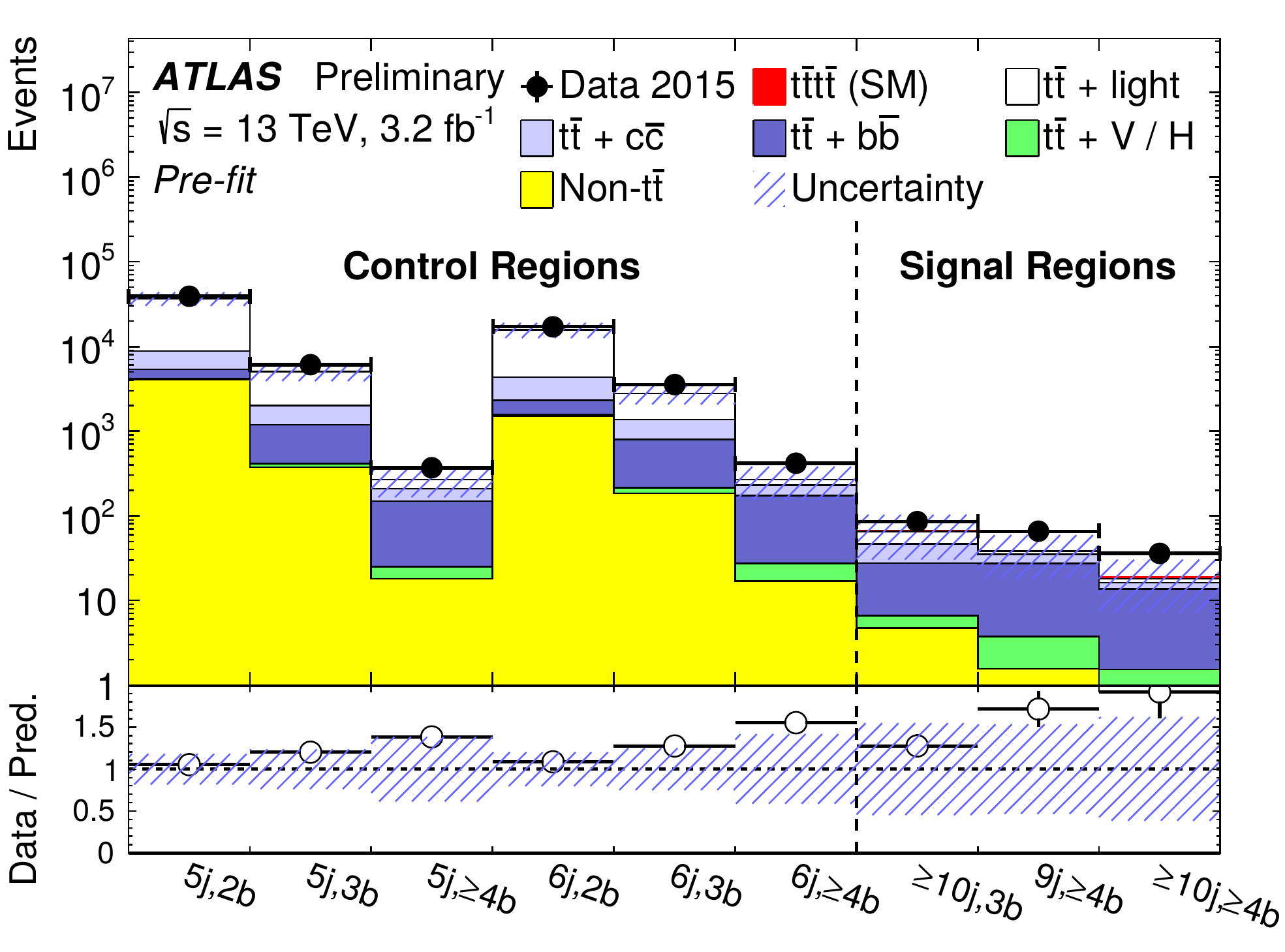}
	\caption{Event yield in the defined measurement regions in the CMS analysis (left), upper limit derivation on $|y_t/y_t^{SM}|$ (center)~\protect\cite{tttt}, and signal and control regions in the ATLAS analysis (right)~\protect\cite{atlas_tttt}.}
	\label{fig:tttt}
\end{figure}

The search for production of four top quarks is performed in events with at least a same-sign lepton pair\cite{tttt} by CMS. Eight measurement regions are built depending on the lepton, jet and b tag multiplicity, and are shown in fig~\ref{fig:tttt}. The signal is extracted by performing a maximum likelihood fit to the measurement region yields, obtaining a \ttbarN\ttbar cross-section of 16.9$^{+13.8}_{-11.4}$ fb, in agreement with SM prediction. This result also allows to constrain the top quark Yukawa coupling, obtaining an upper limit of $|y_t/y_t^{SM}| < 2.1$, as shown in fig~\ref{fig:tttt}.

The ATLAS Collaboration has also performed this measurement using a smaller dataset in the single lepton channel,  in events with large jet multiplicity, obtaining an upper limit on the \ttbarN\ttbar production rate of 21 times the SM value~\cite{atlas_tttt}.

\section{\tW production}
The measurement of tW production is challenging due to the overwhelming presence of \ttbar production in the measurement regions. Both ATLAS~\cite{atlas_tw_inc,atlas_tw_diff} and CMS~\cite{cms_tw} Collaborations have measured this process at $\sqrt{s} = 13$ TeV. ATLAS utilizes the three di-lepton channels (ee, $\mu\mu$ and $e\mu$) together in a smaller dataset, while CMS performs the measurement in the $e\mu$ channel.

\begin{figure}
	\centering
	\includegraphics[height = 4.5cm]{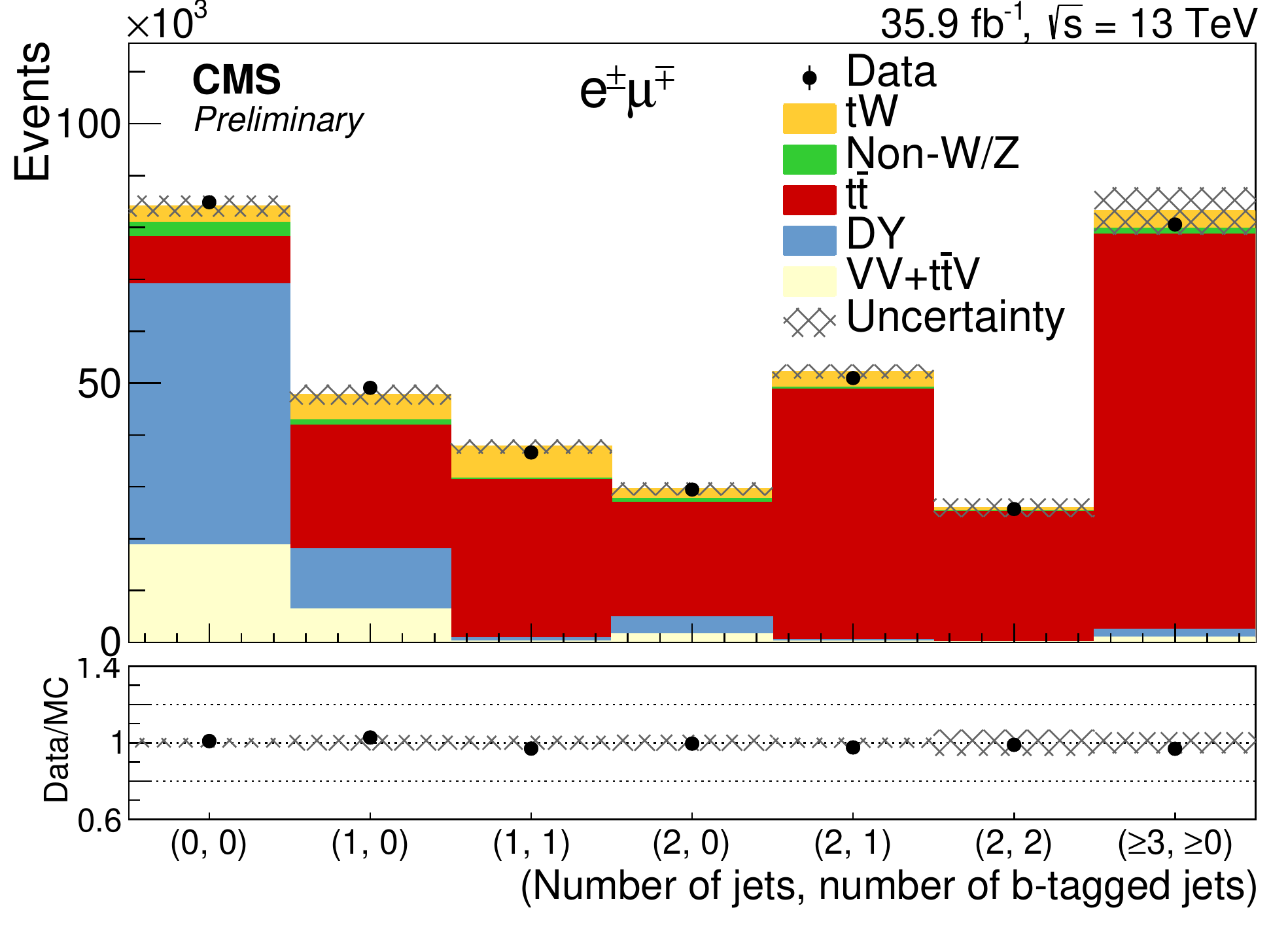}
	\includegraphics[height = 4.5cm]{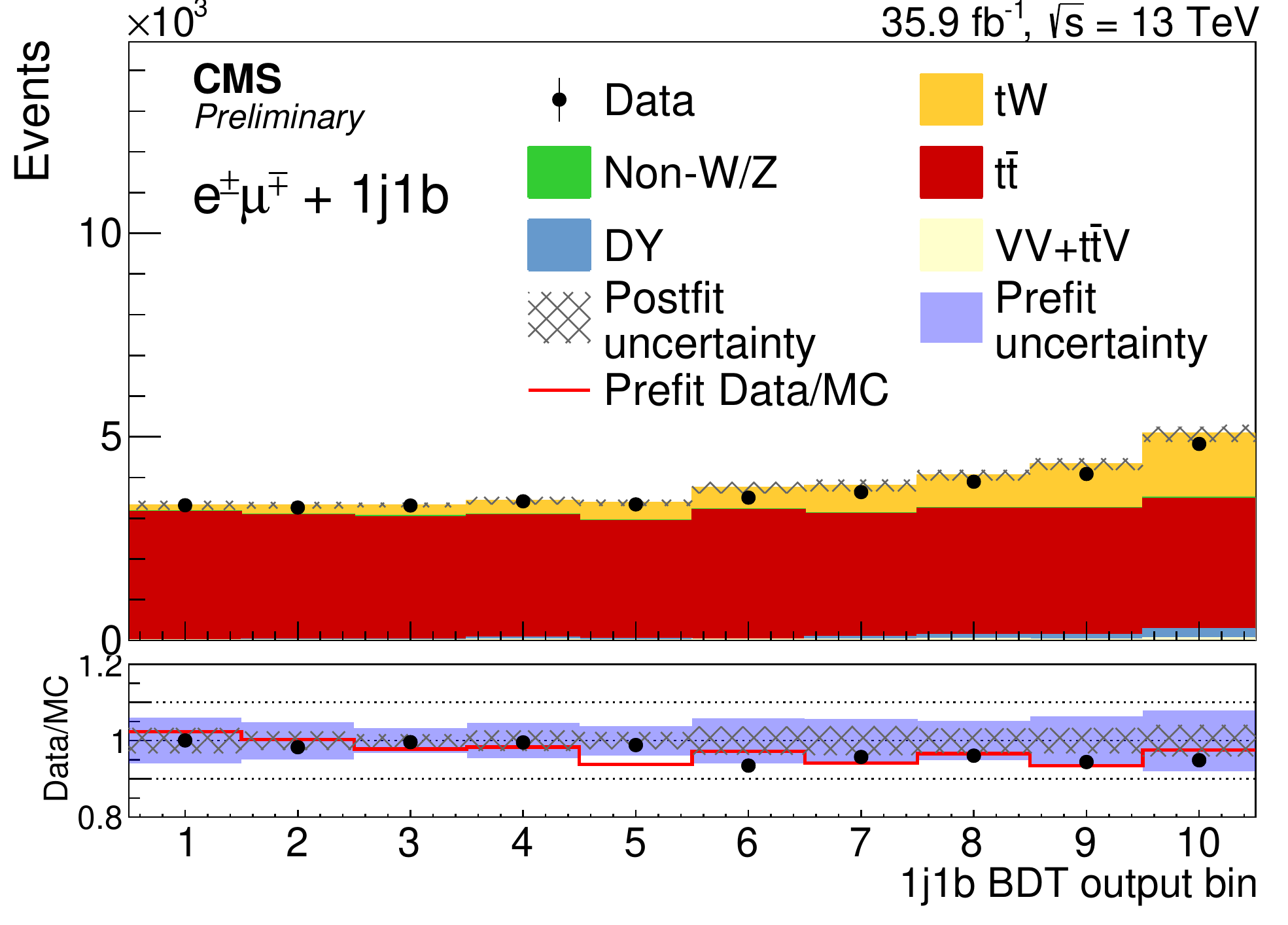}

	\includegraphics[height = 4.5 cm]{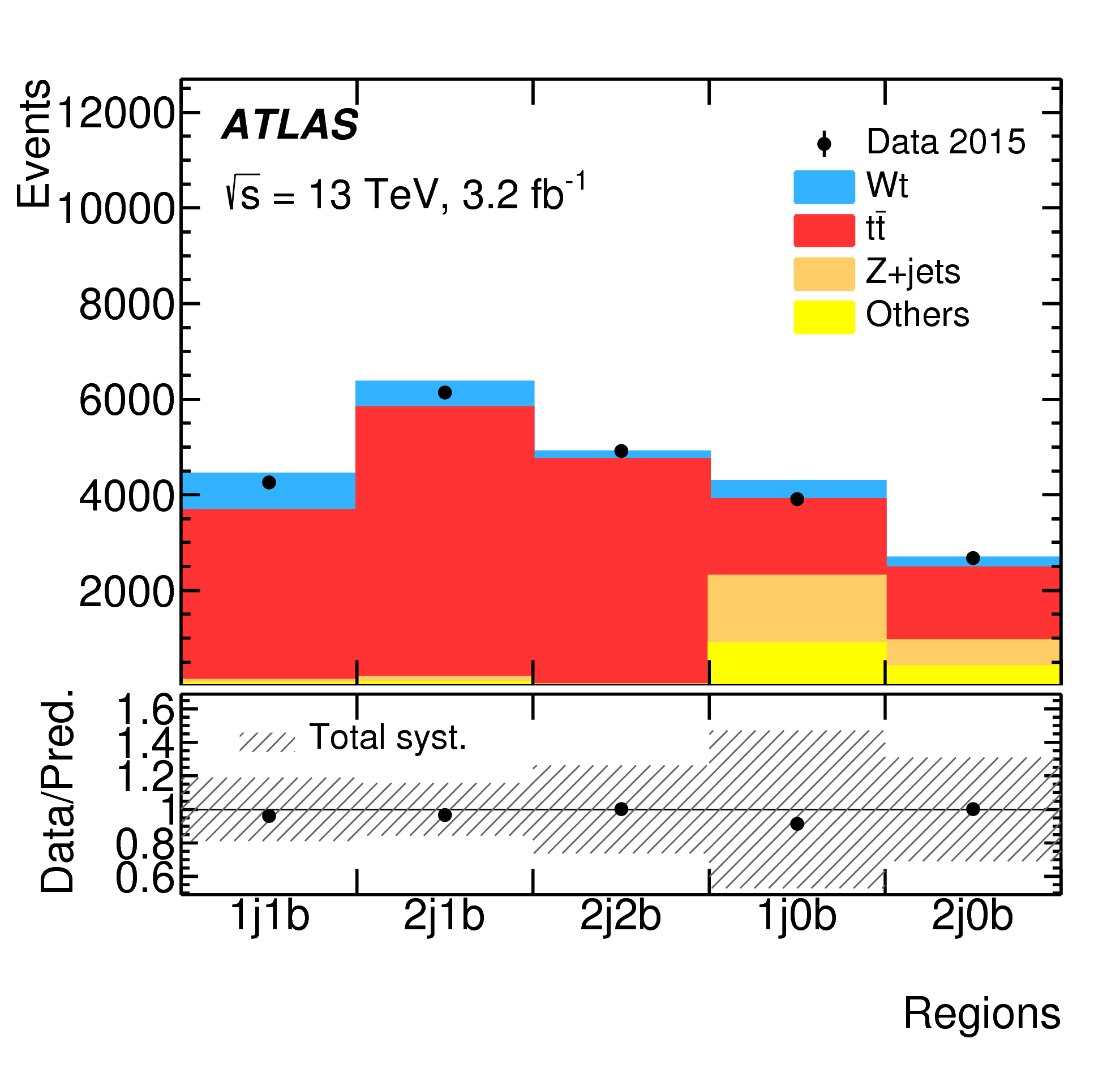}
	\includegraphics[height = 4.5 cm]{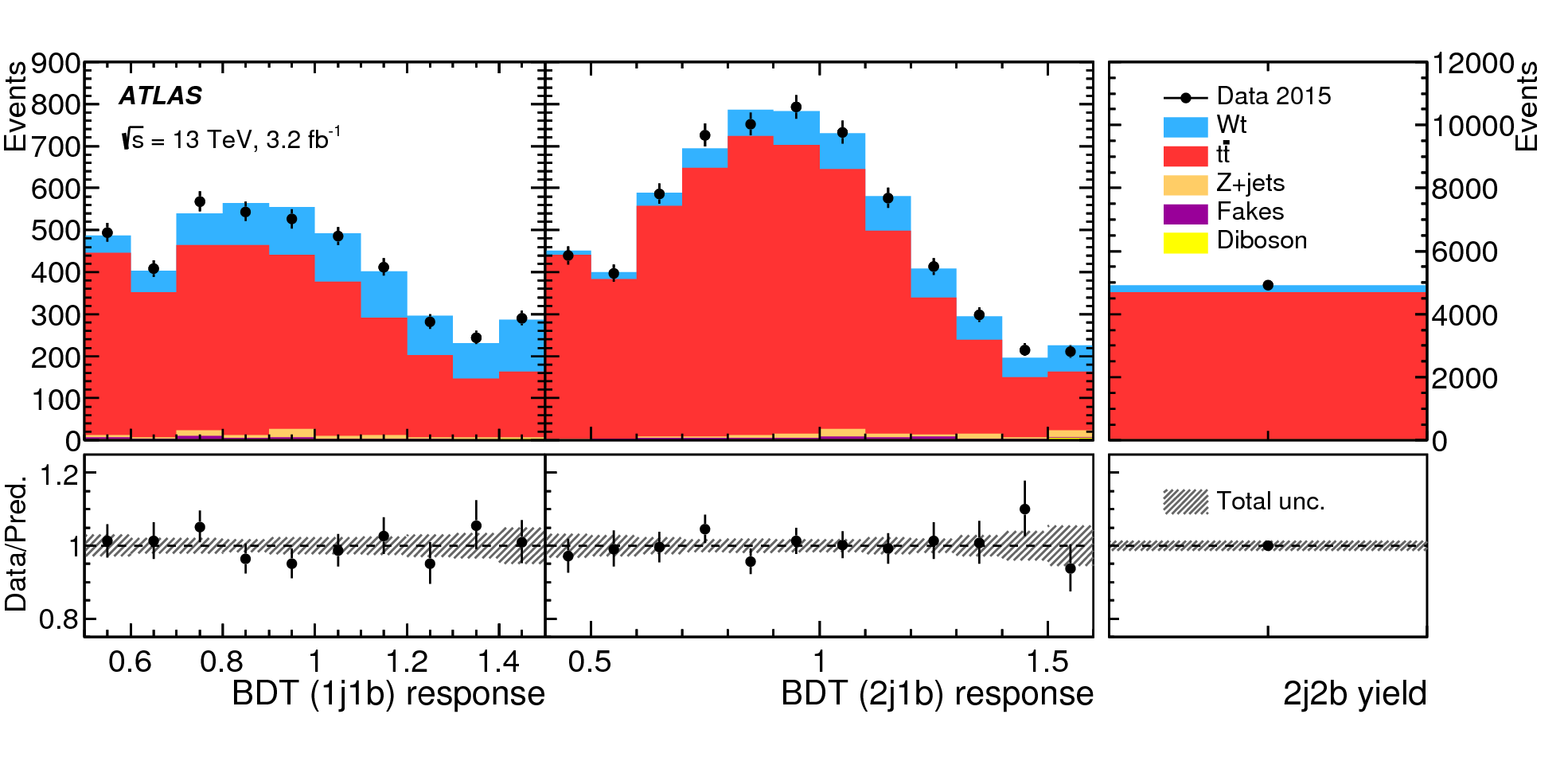}
	\caption{Jet and b-jet multiplicity distribution in selected events (left top and bottom),  MVA discriminator in the 1j1b region (top right) and distributions in the measurement regions (bottom right)~\protect\cite{atlas_tw_inc,cms_tw}.}
	\label{fig:tw_cms}

\end{figure}

Both analyses exploit the different distributions of the jet and b-jet multiplicity between tW and the main background, as shown if fig~\ref{fig:tw_cms}, in order to extract the signal. Both define three measurement regions: 1j1b (exactly one jet that is b-tagged), 2j1b (exactly two jets, out of which one is b-tagged) and 2j2b (exactly two jets that are b-tagged). The first category contains most of the signal, while the second and the third are used to constrain the background.

Both CMS and ATLAS have developed dedicated MVA discriminators for the 1j1b and 2j1b regions in order to discriminate signal from \ttbar, while CMS also uses the subleading jet transverse momentum distribution in the 2j2b region to allow for a better constraint of the systematic uncertainties. The distribution of the MVA discriminators are shown in fig~\ref{fig:tw_cms}.

The signal is extracted by performing a maximum likelihood fit to data in the distribution of the chosen variables, in which systematic uncertainties are parameterized as nuisance parameters.

 The result obtained by the ATLAS Collaboration is a measured inclusive cross-section of $94\ \pm\ 10 \mathrm{\ (stat.)\ }{}^{+28}_{-22}\mathrm{\ (syst.)\ } \pm 2 \mathrm{\ (lumi.)\ pb}$, obtaining a precision of the order of 30\%. CMS obtains a result of $63.1 \pm 3\% \mathrm{ (stat.) } \pm 9\% \mathrm{ (syst.) } \pm 3\% \mathrm{ (lumi) }$ pb, with an accuracy of the order of 10\%. Both results are compatible with the SM prediction and between themselves.

The differential \tW cross-section has also been measured by the ATLAS Collaboration~\cite{atlas_tw_diff}. The measurement is performed in a fiducial region defined by the presence of two leptons and exactly one jet containing b hadrons.  These requirements are applied in the reconstructed-level selection. On top of that, a signal-enhaced region is defined by the MVA discriminator. This approach increases the statistical uncertainty but drastically reduces the \ttbar contribution to the measurement, resulting in an overall reduced uncertainty.

\begin{figure}
	\centering
	\includegraphics[width=0.24\textwidth]{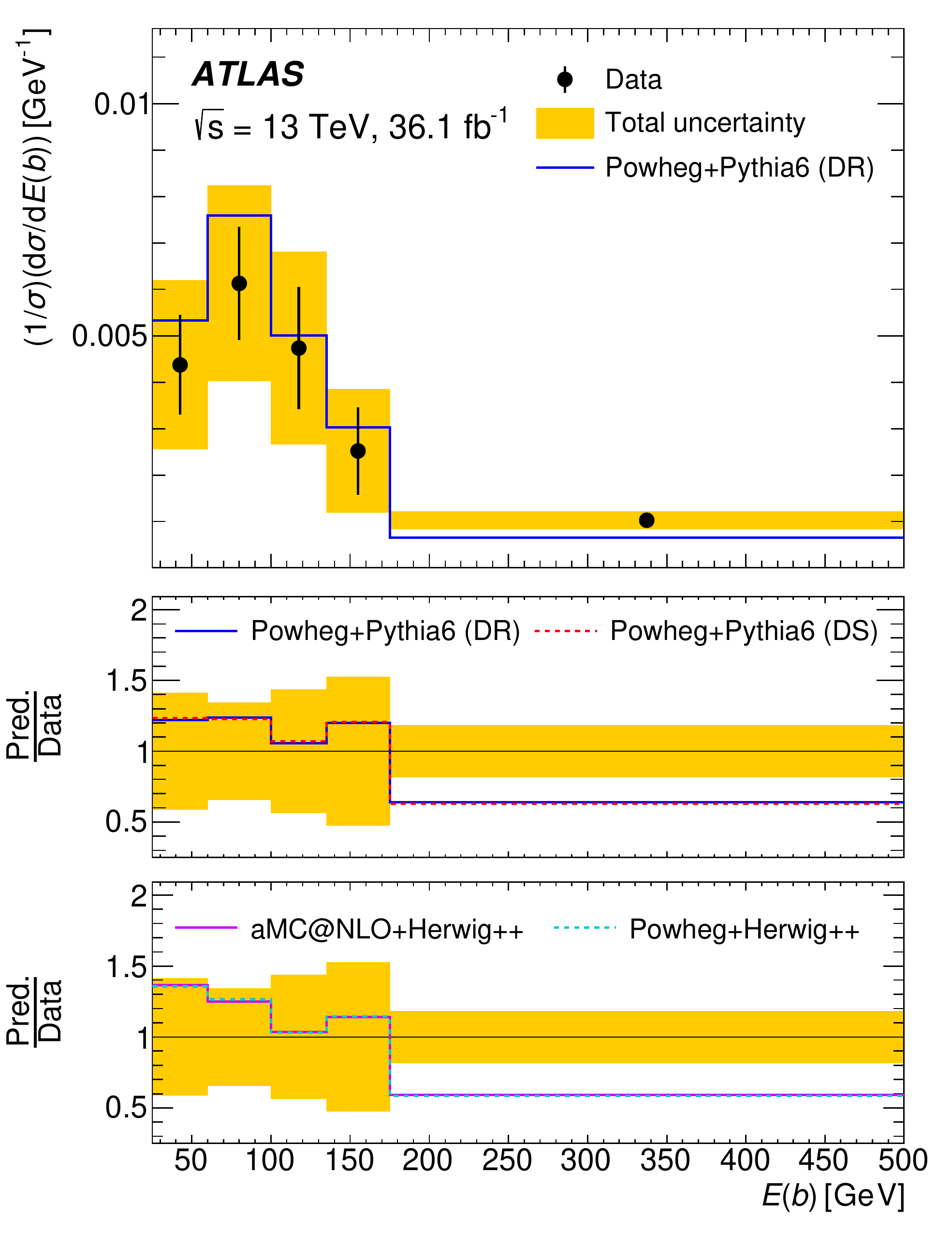}
	\includegraphics[width=0.24\textwidth]{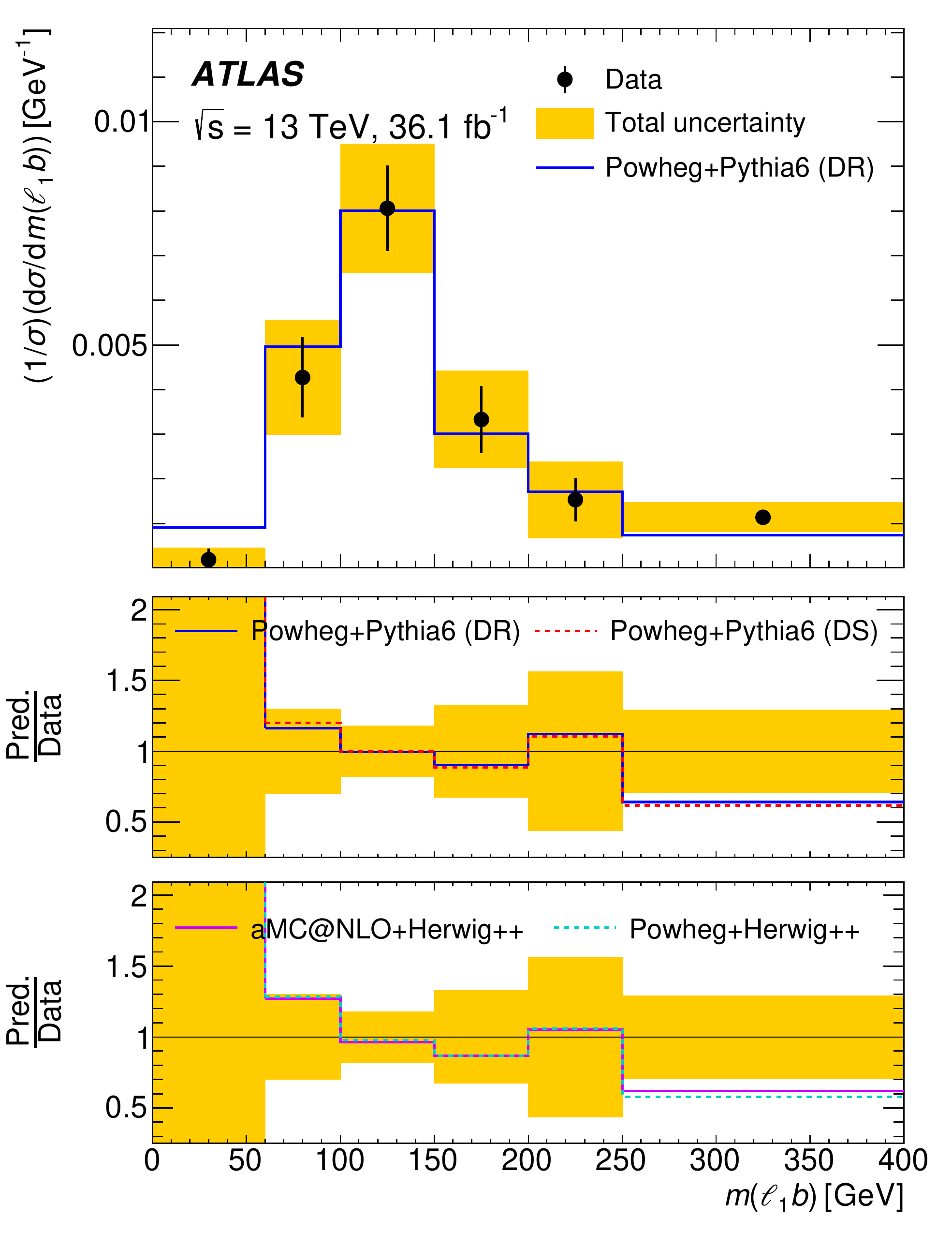}	\caption{Differential cross-section unfolded from data and compared with selected Monte Carlo models~\protect\cite{atlas_tw_diff}.}
	\label{fig:atlas_diff}
\end{figure}

The measurement is performed in six observables that are chosen so they are uncorrelated to the MVA discriminant. The observed distributions are normalized to the measured fiducial cross-section and unfolded to observables based on stable particles produced in Monte Carlo simulations. Figure~\ref{fig:atlas_diff} shows two of the unfolded distributions compared to the results obtained using different Monte Carlo models, showing a fair agreement with all of them.

\section{\tZq production}
ATLAS\cite{atlas_tzq} and CMS\cite{cms_tzq} have shown evidence for \tZq production in pp collisions. Both measurements search for tZq production in events with three leptons, out of which two are candidates of having been produced in the decay of a Z boson.

\begin{figure}
	\centering
	\includegraphics[height = 4.7cm]{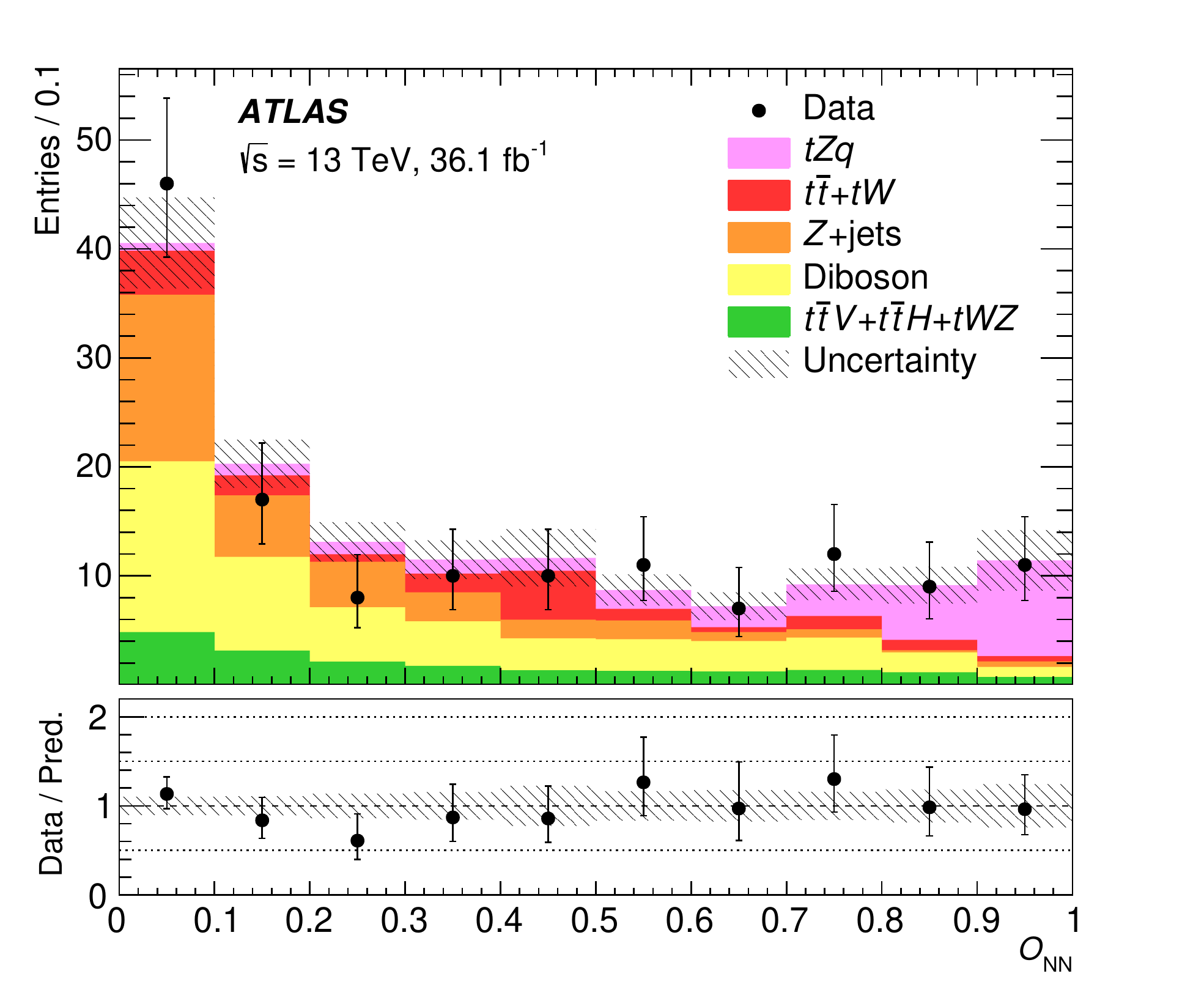}
	\includegraphics[height = 4.7cm]{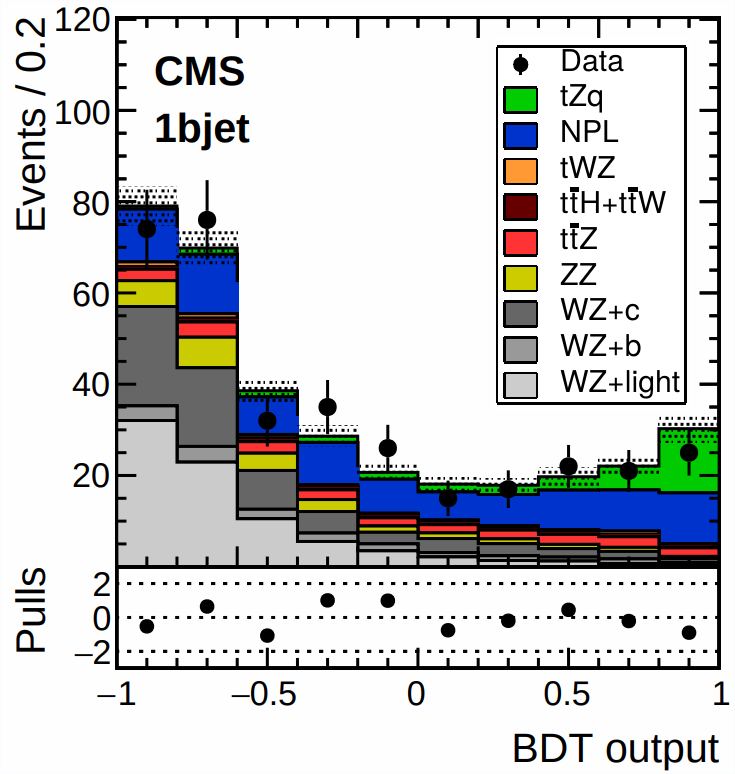}
	\caption{Distribution of MVA discriminators used for the signal extraction in the most sensitive regions, after the signal extraction fit is performed~\protect\cite{atlas_tzq,cms_tzq}.}
	\label{fig:tzq_atlas_and_cms}
\end{figure}

The measurement by ATLAS is performed in events with exactly 2 jets, out of which one of them must be b-tagged. The remaining jet typically recoils against the tZ system and can be emitted more colinearly with respect to the beam direction. The signal is extracted by performing a maximum likelihood fit to the distribution of a MVA discriminator trained to disentangle signal from the backgrounds, which is shown in figure~\ref{fig:tzq_atlas_and_cms}.

The main source of background in this measurement is due to events with non-prompt leptons, and are estimated in a twofold approach: events with a Z boson and a non-prompt lepton are estimated using a data-driven method, while the remaining components of this background -mainly \ttbar- are estimated from simulations and validated in dedicated control regions.

The CMS analysis utilizes three measurement regions defined by the number of b-tagged jets and jets. The 1bjet region is defined as events containing two or three jets, one tagged a b jet, and is the most signal enriched one. The 2bjets and 0bjets regions are used to control the \ttZ and WZ backgrounds, repectively. The former is defined as events with at least two jets, with at least two b-tagged, while the latter is formed with events with at least one jet, but zero b jets.

Signal is extracted by performing a maximum likelihood fit to the distributions of two MVA discriminants in the 1bjet and 2bjet regions, while the transverse mass distribution is employed in the 0bjet region. The distribution of the MVAs is shown in figure~\ref{fig:tzq_atlas_and_cms}.

The result by ATLAS reports a signal strengh of $\mu = 0.75\ \pm\ 0.21\ \mathrm{ (stat.) } \pm\ 0.17\ \mathrm{ (syst.) }\ \pm\ 0.05\ \mathrm{ (th.) }$, while the result by CMS obtains a signal strength of $\mu = 1.31^{+0.35}_{-0.33}\ \mathrm{ (stat.) }\ {}^{+0.31}_{-0.25}\ \mathrm{ (syst.) } $, both consistent with the SM predictions.

\section*{Acknowledgments}

The author wishes to thank the Becas Severo Ochoa del Principado de Asturias for partially funding this work.

\end{document}